\documentclass[twocolumn,twoside,nofootinbib]{revtex4}
\usepackage{graphicx}
\usepackage[utf8]{inputenc}
\usepackage{amsmath,latexsym}
\usepackage{fancyhdr}
\begin{document}

\title{Lepton non-universality in $B$-decays in the minimal leptoquark gauge model}

%

\author{Michal Malinsk\'y}
\affiliation{Institute of Particle and Nuclear Physics, Charles University, V Hole\v{s}ovi\v{c}k\'ach 2, 18 000 Prague, Czech Republic}

\begin{abstract}
The anomalies in semileptonic $B$-decays are often attributed to new physics scenarios featuring leptoquark degrees of freedom. Attempts to accommodate all the deviations at once usually result in an elaborate model building way beyond the minimal scenarios containing the desired degrees of freedom. However, as it is far from clear whether all these signals (if any) survives the future experimental scrutiny, in this contribution we decided to take the opposite standpoint and review the actual room available for various $B$-decay anomalies within the very minimal low-energy leptoquark gauge model based on the $SU(4)_{C}\otimes SU(2)_{L}\otimes U(1)_{R}$ symmetry group.   
\end{abstract}

\maketitle

\thispagestyle{fancy}


\section{Introduction}
The observed anomalies in the semileptonic $B$-decays, in particular those covered by the $R_{K^{(*)}}$ and $R_{D^{(*)}}$ parameters 
\begin{eqnarray}
\label{RK}
R_{K^{(*)}}  &=& \frac{\Gamma(\bar B \to \bar K^{(*)} \mu^+ \mu^-)}
                   {\Gamma(\bar B \to \bar K^{(*)}e^+ e^-)}\\
R_{D^{(*)}} &=& \frac{\Gamma(\bar B \to D^{(*)} \tau \bar \nu)}
                   {\Gamma(\bar B \to D^{(*)} l \bar \nu)}
      \hspace{5mm}(l=e,\mu) \,,  
\end{eqnarray}
as reported by Belle \cite{Matyja:2007kt,Bozek:2010xy,Huschle:2015rga} and LHCb \cite{Aaij:2014ora,Aaij:2015yra,Aaij:2017vbb}, are among the hottest topics in particle phenomenology nowadays; as potential harbingers of physics beyond the Standard Model (SM) they receive enormous attention not only by the experimentalists but also by the theoretical community. 

In addressing them, both the bottom-up and top-down approaches are generally entertained. As for the former which, as usual, prevailed within the initial attempts, there has been a myriad of hypothetical new fields of different masses, spins and couplings added to the SM and their impact on (not only) the observables above has been studied. 

Among the most remarkable results of this first-phase accounts there was the observation \cite{Buttazzo:2017ixm,Angelescu:2018tyl} that a vector leptoquark field (called $U_{1}$ by convention set in~\cite{Dorsner:2016wpm}) with the SM quantum numbers $(3,1,+\frac{2}{3})+h.c.$ is particularly suitable for the job if its mass and couplings are adjusted to a certain narrow range. 

As intriguing as this option is it becomes rather problematic when a UV-complete model including this type of a leptoquark within a consistent gauge framework is eventually sought for. The first guess, namely, the $SU(4)_{C}\otimes SU(2)_{L}\otimes U(1)_{R}$ gauge scheme which, among other things, represents the minimal gauge scenario involving $U_{1}$, fails spectacularly; the main reason is that the desired coupling of  the $U_{1}$ to the matter currents is way off the one suggested by the embedding of the $SU(3)_{c}\otimes U(1)_{B-L}$ within the unified $SU(4)_{C}$. As a result, there is a number of studies (e.g., \cite{DiLuzio:2017vat,Calibbi:2017qbu}) in which a favourable pattern of the $U_{1}$ couplings has been attained for the usual price of a more complicated gauge group structure.

The generic baroqueness of the potentially realistic versions of such scenarios, together with the significant shifts in some of the recent  data  updates (see, e.g., \cite{Aaij:2019wad,Prim:2019hyn} and references therein) towards the SM expectations, makes it natural to speculate that, maybe, not all the aforementioned anomalies are real and that the model-building issues may be emerging because of a mere overloading of the theoretical schemes at stakes. 

From this perspective, a careful reassessment of the potential room within the most minimal models for various subsets of the entire load of the known anomalies may be of some interest, the more that this is a very natural path to follow in the top-down approach in general.

\section{$B$-anomalies in the minimal gauge model with leptoquarks}
In this presentation we shall stick to the most minimal gauge extension of the Standard Model featuring leptoquark degrees of freedom which is based on the $SU(4)_{C}\otimes SU(2)_{L}\otimes U(1)_{R}$ gauge symmetry \cite{Smirnov:1995jq,Perez:2013osa}. Apart from the gauge sector (including -- besides the SM gauge fields -- the $U_{1}$ leptoquark) and the matter fields residing (each generation) in just three irreducible representations $F\equiv (4,2,0)$, $f^{c}_{u}\equiv (\bar 4,1,-\tfrac{1}{2})$ and $f^{c}_{d}\equiv (\bar 4,1,+\tfrac{1}{2})$ it contains a number of additional scalar degrees of freedom emanating from the Higgs sector responsible for the $SU(4)_{C}\otimes SU(2)_{L}\otimes U(1)_{R}\to SU(3)_{c}\otimes SU(2)_{L}\otimes U(1)_{Y}$ symmetry breaking. Among these, the pair of scalar leptoquarks $R_{2}\equiv (3,2,+\tfrac{7}{6})$ and  $\tilde R_{2}\equiv (\bar 3,2,-\tfrac{1}{6})$ will play a prominent role in what follows. For more details about the structure of the model under consideration the reader is deferred to the original studies~\cite{Faber:2018qon,Faber:2018afz}.

Unlike for the vector $U_{1}$ whose mass in the minimal model is constrained to be above about $80$~TeV (by, namely, the lepton-flavour-violation (LFV) in meson decays such as $K_{L}\to \mu e$, $B\to \ell \ell'$ etc.) \cite{Valencia:1994cj,Smirnov:2018ske}, which disqualifies it from the role of a mediator behind $R_{K^{(*)}}$ and $R_{D^{(*)}}$ even for the maximal coupling allowed by the $SU(4)_{C}\otimes SU(2)_{L}\otimes U(1)_{R}$ gauge symmetry, the interactions and masses of the scalars\footnote{Focusing on the scalar leptoquarks only it is very natural to ask why we do not simply drop the entire extra gauge structure and work with just the SM + the desired degrees of freedom. The point here is that in doing so we would deprive ourselves from all the links between their Yukawa couplings and the Yukawas of the Higgs doublet, aka the mass matrices of the SM fermions that we do know something about; hence, the narrative would drift back to the bottom-up approach.} $R_{2}$ and $\tilde R_{2}$ are subject to a way milder constraints.  As for the former, their Yukawa couplings are tightly correlated to the mass matrices of the SM fermions, while the latter are typically subject to sum-rules deriving from the minimization of the relevant scalar potential such as (see~\cite{Faber:2018qon}): 
\begin{equation}
m^{2}_{R_{2}}+m^{2}_{\tilde R_{2}}\sim \frac{3}{2}\left(m_{G}^{2}+2 m_{H}^{2}\sin^{2}\beta\right)\,;
\end{equation}
here $G$ and $H$ denote the extra scalar $SU(2)_{L}$-doublets ($SU(3)_{c}$  octet and singlet, respectively) 
present in the spectrum of the model and $\beta$ is the mixing angle among the two colourless $SU(2)_{L}$ doublets of the model. 
The only constraint here is that at least one of the masses (on each side) should be at around the symmetry-breaking scale defined by the mass of $U_{1}$ and, thus, in the 100~TeV ballpark. This, among other  things\footnote{Another interesting scenario, especially from the collider perspective, emerges if the scalar gluons $G$ happen to be light, see the discussion in \cite{Faber:2018afz}.}, admits for an attractive option that one (and only one) of $R_{2}$ or $\tilde R_{2}$ may be as light as few TeV, thus having all the prerequisites to exhibit itself in some way in the low-energy physics such as the $B$-decay anomalies.

Even at this initial point, several interesting observations regarding the relevant phenomenology can already be made:
i) As only one of the $R_{2}$ and $\tilde R_{2}$ leptoquarks may be light it is impossible to address the $R_{D^{(*)}}$ anomalies within the minimal $SU(4)_{C}\otimes SU(2)_{L}\otimes U(1)_{R}$ gauge extensions of the SM, see, e.g., \cite{Faber:2018qon} and references therein. One is, thus, mostly left with the semileptonic $B$-decays into Kaons, i.e., $R_{K^{(*)}}$.
ii) Since the light $\tilde R_{2}$ option is strongly disfavoured \cite{Angelescu:2018tyl,Becirevic:2015asa} (as it leads to $R_{K^{*}}>1$, at odds with the observation, and in principle even to the issues with nucleon decay), the light  $R_{2}$ remains as the only potentially viable option.

Needless to say, the case of a light $R_{2}$ has been thoroughly studied from the bottom-up perspective so, at first glance, there seems to be little room for adding anything new at this point. Interestingly, a closer inspection reveals an implicit assumption imposed in most of these cases which has to do with the apparent suppression of the muonic rates with respect to the SM ones which looks like a  universal cure to most of the anomalies. Since, however, we are not aiming at accommodating all of these at once, an alternative scenario in which the first rather than the second generation leptoquark couplings are affected is naturally brought back into play; in other words, $R_{K^{(*)}}<1$ may result from the enhancement of the denominator in (\ref{RK}) rather than the usual suppression of the numerator.
\subsection{Phenomenology of a light $R_{2}$ scalar}
The interactions of the two $R_{2}$ charged components within the minimal $SU(4)_{C}\otimes SU(2)_{L}\otimes U(1)_{R}$ gauge model are driven by the lagrangian of the form
\begin{equation}
\mathcal{L}_\mathrm{R_2} 
=\overline{\hat d_L}\,  {Y}_4^{de} \, \hat e_R \, R_2^{+2/3}
+ \overline{\hat u_L}\, V_\mathrm{CKM}{Y}_4^{de} \, \hat e_R \, R_2^{+5/3} 
+ h.c.
\end{equation}
where the hatted matter fields correspond to the mass eigenstates, $V_{CKM}$ is the Cabibbo-Kobayashi-Maskawa matrix of the SM and $Y_{4}^{de}$ is a matrix of the relevant Yukawa couplings; this, due to the extended gauge symmetry, is tightly correlated to the fermionic masses: 
\begin{equation}
Y_4^{de}=
\sqrt{\frac{3}{2}} \frac{1}{v \cos\beta} 
\left(D_{d} U_d - V_d D_e \right)\,.
\label{YdeSU4}
\end{equation}
Here $v$ stands for the electroweak VEV, $D_{u,d}$ denote the diagonal mass matrices of the up and down quarks and $U_{d}$ and $V_{d}$ stand for the left and right diagonalization rotations in the down-quark sector, see~\cite{Faber:2018afz}. 

As a matter of fact, there is a simple way how to translate most of the phenomenological constraints on the market into the constraints on the elements of the $3\times 3$ matrix $Y_{4}^{de}$:
i) The stringent constraints on the LFV kaon and $B$-meson decays boil down to limits onto various products of its first and second column entries. In a self-explanatory notation, smallness of $y_{se}.y_{d\mu}$, $y_{se}.y_{d\mu}$ and so on is demanded.
ii) In a similar fashion, the pure LFV constrains, in particular those from $\mu\to e\gamma$, $\mu\to 3e$, can be accommodated if the products $y_{se}.y_{s\mu}$ and  $y_{be}.y_{b\mu}$ are strongly suppressed.
iii) The $y_{de}$ element is constrained by, e.g.,  the atomic parity violation limits~\cite{Dorsner:2016wpm}.

With all the constraints above the desired enhancement of the numerator in~(\ref{RK}) is achieved only if $y_{se}.y_{be}$ is non-negligible. 
Hence, one immediately arrives at a very specific texture of $Y_{4}^{de}$ that may conform all the existing SM limits and yet provide large-enough effects in $R_{K^{(*)}}$ in the schematic form
\begin{equation}
\label{texture}
Y_{4}^{de}\sim 
\left(
\begin{array}{ccc}
\cdot & \cdot&  \Box\\
\bullet & \cdot &  \Box\\
\bullet& \cdot & \Box
\end{array}
\right)\,,
\end{equation}
where the dots ($\cdot$) represent entries which are supposed to be strongly suppressed, the bullets ($\bullet$) stand for entries that should be non-negligible and the boxes ($ \Box$) remain to a large extent undetermined.

\subsection{$R_{2}$ leptoquark Yukawa textures in the minimal $SU(4)_{C}\otimes SU(2)_{L}\otimes U(1)_{R}$ model}

The critical question to answer is obviously whether a specific texture such as (\ref{texture}) is attainable within the minimal model formula~(\ref{YdeSU4}) and if, indeed, the resulting phenomenology is even quantitatively compatible with the SM phenomenology provided  $R_{K}$ and $R_{K^{*}}$ were taken at their face values. As this has been studied thoroughly in a pair of articles~\cite{Faber:2018qon,Faber:2018afz} here we shall only stick with the results of the extensive numerical simulations therein:
i) It has been shown in \cite{Faber:2018qon} that in the case of symmetric Yukawa couplings (which is a scenario motivated by the possible embedding of the $SU(4)_{C}\otimes SU(2)_{L}\otimes U(1)_{R}$ gauge structure into the $SO(10)$ GUT) there is no way to entirely suppress the 2nd (muonic) column of  $Y_{4}^{de}$ due to the strong correlation between the left- and right-handed rotations $V_{d}$ and $U_{d}$ in (\ref{YdeSU4}). This typically exhibits itself in overly large branching ratios for $K_{L}\to \mu e$ and/or $\mu\to e\gamma$ etc.
ii) In the case of a fully general Yukawa structure~\cite{Faber:2018afz} the muonic column of $Y_{4}^{de}$ may be entirely zeroed out if $U_{d}$ and $V_{d}$ are chosen carefully. At the same time, the residual freedom in their selection is rather small (limited to a single angle $\phi$ in the real Yukawa case) which, among other  things, makes it impossible to suppress more than a single entry in the third (tau) column of  $Y_{4}^{de}$ at a time.

Thus, the LFV tau decays (with predicted upper limits of $\text{BR}(\tau \to e\ell^{+}\ell^{-})\leq 2.7 \times 10^{-8}$, $\text{BR}(\tau \to e\gamma)\leq 3.3 \times 10^{-8}$, $\text{BR}(\tau \to e\pi^{0})\leq 8 \times 10^{-8}$ and $\text{BR}(\tau \to e\pi^{+}\pi^{-})\leq 2.2 \times 10^{-8}$) provide a clear smoking-gun signal of the minimal  $SU(4)_{C}\otimes SU(2)_{L}\otimes U(1)_{R}$ gauge model account for the observed $R_{K}$ and $R_{K^{*}}$ anomalies in the $B$-meson decays. 

\section{Conclusions}
In this contribution we attempted to map the room for accommodating a selection of the anomalies in the $B$-meson decays observed recently by Belle and LHCb in the context of the minimal gauge extension of the Standard Model featuring leptoquark degrees of freedom. Its rigid  $SU(4)_{C}\otimes SU(2)_{L}\otimes U(1)_{R}$ gauge structure admits only one of its scalar leptoquarks, namely, the one with quantum numbers $(3,2,+\tfrac{7}{6})$, to be light enough and, at the same time, acquire a suitable pattern of the Yukawa couplings in order to generate large-enough effects in the quantities of interest (mainly $R_{K}$ and $R_{K^{*}}$) while still maintaining the SM compatibility in all other channels. Remarkably enough, the relevant dynamics leaves behind traces in the LFV $\tau$-decay channels at the level that may be tested in the upcoming runs of the existing experiments such as Belle II.  

\begin{acknowledgments}
The author thanks the organizers of the FPCP2019 conference for all the hospitality and care with which the meeting was put together. His participation and work has been supported by the Grant Agency of the Czech Republic, Project No. 17-04902S. 
\end{acknowledgments}

\bigskip 

\end{document}